\documentclass[preprint,preprintnumbers, prd, floatfix, superscriptaddress,nofootinbib] {revtex4-1}
\usepackage{epsfig}
\usepackage{subfigure}
\usepackage{dcolumn}% Align table columns on decimal point
\usepackage{bm}% bold math
\usepackage[usenames ,dvipsnames]{xcolor}
\usepackage{slashed}
\usepackage{graphicx,color}
\usepackage{hyperref}  
\begin{document}
	
\title{\boldmath Improvement of $q^2$ resolution in semileptonic decays based on machine learning}

\author{Panting Ge}
\affiliation{School of Physics and Technology, Wuhan University, Wuhan 430072, China}

\author{Xiaotao Huang}
\affiliation{The Institute for Advanced Studies, Wuhan University, Wuhan 430072, China}

\author{Miroslav Saur}
\affiliation{Technische Universit\"{a}t Dortmund, Dortmund 44227, Germany}

\author{Liang Sun}
\email{Corresponding author: sunl@whu.edu.cn}
\affiliation{School of Physics and Technology, Wuhan University, Wuhan 430072, China}

\date{Received: date / Accepted: date}

\begin{abstract}
    The neutrino closure method is often used to obtain kinematics of semileptonic decays with one unreconstructed particle in hadron collider experiments.
    The kinematics of decays can be deducted by a two-fold ambiguity with a quadratic equation.
	To resolve the two-fold ambiguity, a novel method based on Machine Learning (ML) is proposed.
	We study the effect of different sets of features and regressors on the improvement of reconstructed invariant mass squared of $\ell \nu$ system~($q^2$).
	The result shows that the best performance is obtained by using the flight vector as the features, and the multilayer perceptron (MLP) model as the regressor. 
    Compared with the random choice, the MLP model improves the resolution of reconstructed $q^2$ by $\sim$40\%.
    Furthermore, the possibility of using this method on various semileptonic decays is shown.
\end{abstract}

\maketitle

\section{Introduction}
\label{sec:introduction}
Semileptonic decays, mediated by a virtual $W$ boson which produces one lepton and the corresponding neutrino in addition to one or more hadrons, offer a good platform to study the weak as well as strong interaction effects~\cite{Dingfelder:2016twb}. 
Studies of semileptonic decays, therefore, have been paid much more attention in recent years, especially for the purposes of precise measurements on the Cabibbo-Kobayashi-Maskawa (CKM) matrix elements~\cite{Charles:2004jd,Ricciardi:2019zph}, such as the determination of $\vert V_{ub} \vert$ and $\vert V_{cb} \vert$.  
The precision measurement of the CKM matrix elements help predict other branching fractions, such as $B \to \tau \nu_\tau$.
Additionally, the recent measurements of the branching fraction ratios $R(D^\star) = {\cal B}(B\to D^{(\star)}\tau\nu)/{\cal B}(B\to D^{(\star)}\mu\nu)$ measured in experiments show a slight disagreement with the Standard Model predictions~\cite{HFLAV:2016hnz}. 
Based on the above, the studies of semileptonic decays by LHCb experiment, which focuses on a heavy-flavour studies in a forward region, show an increasing trend, although the presence of an unreconstructed neutrino is experimentally challenging.

At $B$-factories operating at the $\mathbf{\Upsilon}$(4S) resonance, the kinematics of missing particles in $B$ mesons can be reconstructed by balancing against the $\bar{B}$ decay~\cite{Ciezarek:2016lqu},  
while in hadron collider experiments the studies of semileptonic decays pose a technical challenge~\cite{Gambino:2020jvv} due to the unreconstructed neutrino in the final state. 
First of all, a large Lorentz boost can be produced by hadron collider experiments, especially at the forward rapidity covered by the LHCb experiment~\cite{Alves:2008zz}, which is one of the major experiments at LHC.
Secondly, the decay kinematics can be restricted by the $b$-hadron decay vertex and the measured flight vector which connects with the primary $pp$ interaction vertex~\cite{Dambach:2006ha}.
Finally, the mass of single missing particles can be deduced from the conservation of four-momentum. 
Conservation of the transverse momentum to the flight vector provides two independent constraints on the semileptonic decays as well.
A third constraint is the parent $b$-hadron mass should be conserved, though this condition has an ambiguity which produces two solutions. 

A recently proposed lattice QCD method~\cite{Detmold:2015aaa} for the precise calculation of the relevant hadronic form factors shows that the magnitudes of the CKM matrix elements can be calculated based on these known form factors and measurements of $\Lambda_b^0 \to p\mu\nu$ and $\Lambda_b^0 \to \Lambda_c\mu\nu$.
At the same time, a measurement of the ratio $\rm \frac{\vert V_{ub} \vert}{\vert V_{us} \vert}$ with a newly observed exclusive decay $\Lambda_b^0 \to p\mu\nu$ and $\Lambda_b^0 \to \Lambda_c\mu\nu$ has been performed by LHCb experiment~\cite{LHCb:2015eia}.
This measurement has a significant effect on global fits to the parameters of the CKM matrix. 
Similarly, the single most precise determination of $\rm \frac{\vert V_{ub} \vert}{\vert V_{cb} \vert}$ has been obtained from a 2+1-flavour lattice QCD calculation with domain-wall light quarks and relativistic heavy quarks, which is based on the mentioned $B_s^0$ decay mode $B_s^0 \to K\mu\nu$~\cite{Flynn:2015mha}.
LHCb recently made the first observation of the suppressed semileptonic decay $B_s^0 \to K\mu\nu$, and subsequently measured the ratio of the CKM matrix elements $\rm \frac{\vert V_{ub} \vert}{\vert V_{cb} \vert}$ at low and high $B_s^0 \to K\mu\nu$ momentum transfer~\cite{Hicheur:2764845}.
One of the challenges for the determination of CKM matrix elements in hadron collider experiments is to infer $q^2$. 
To calculate the above, we need to reconstruct the neutrino momentum with a reasonable precision. 

In the Ref.~\cite{Ciezarek:2016lqu}, a linear regression based on estimating of the $b$-hadron momentum, using flight vector as input, can then be used to resolve the quadratic ambiguity.
Based on the above study, we proposed a method using the MLP regressor based on 0.54 of the correlation coefficient of $\rm 1/sin(\theta_{\rm flight})$ versus the $b$-hadron momentum. 
This implies that there is underlying non-linear dependence of the target on features which can not be captured by linear regressor.
The work presented below aims to improve the $q^2$ resolution of semileptonic decays in hadron collider environment, based on ML with the Python library {\tt scikit-learn}~\cite{Pedregosa:2011ork}.
At first, the formula for the decay kinematics with a missing particle is briefly introduced.
Then for this study, simulated events based on the {\tt RapidSim} fast Monte Carlo (MC) generator~\cite{Cowan:2016tnm} are used to simulate semileptonic decays in $pp$ collision.
Furthermore, different sets of features and regressors have been studied to select the flight vector and MLP model with the best performance. 
Then, using the decay $B_s^0 \to K\mu\nu$ as a test channel, the resolution improvement of $q^2$ is compared with random choice and the linear regressor method introduced in Ref.~\cite{Ciezarek:2016lqu}. 
Finally, in order to examine the performance and to obtain a credible conclusion, 
other semileptonic decay channels are tested as well.
This paper will use LHCb as a model detector, but the ideas should be available to any other hadron collider experiment in the future.

\section{Theoretical derivation of neutrino momentum}
\label{sec:theorecticalDerivation}
\begin{figure} % figure 1
	\begin{minipage}{\columnwidth}
		\centering
		\includegraphics[width=.8\textwidth,clip,trim=0 0 0 0]{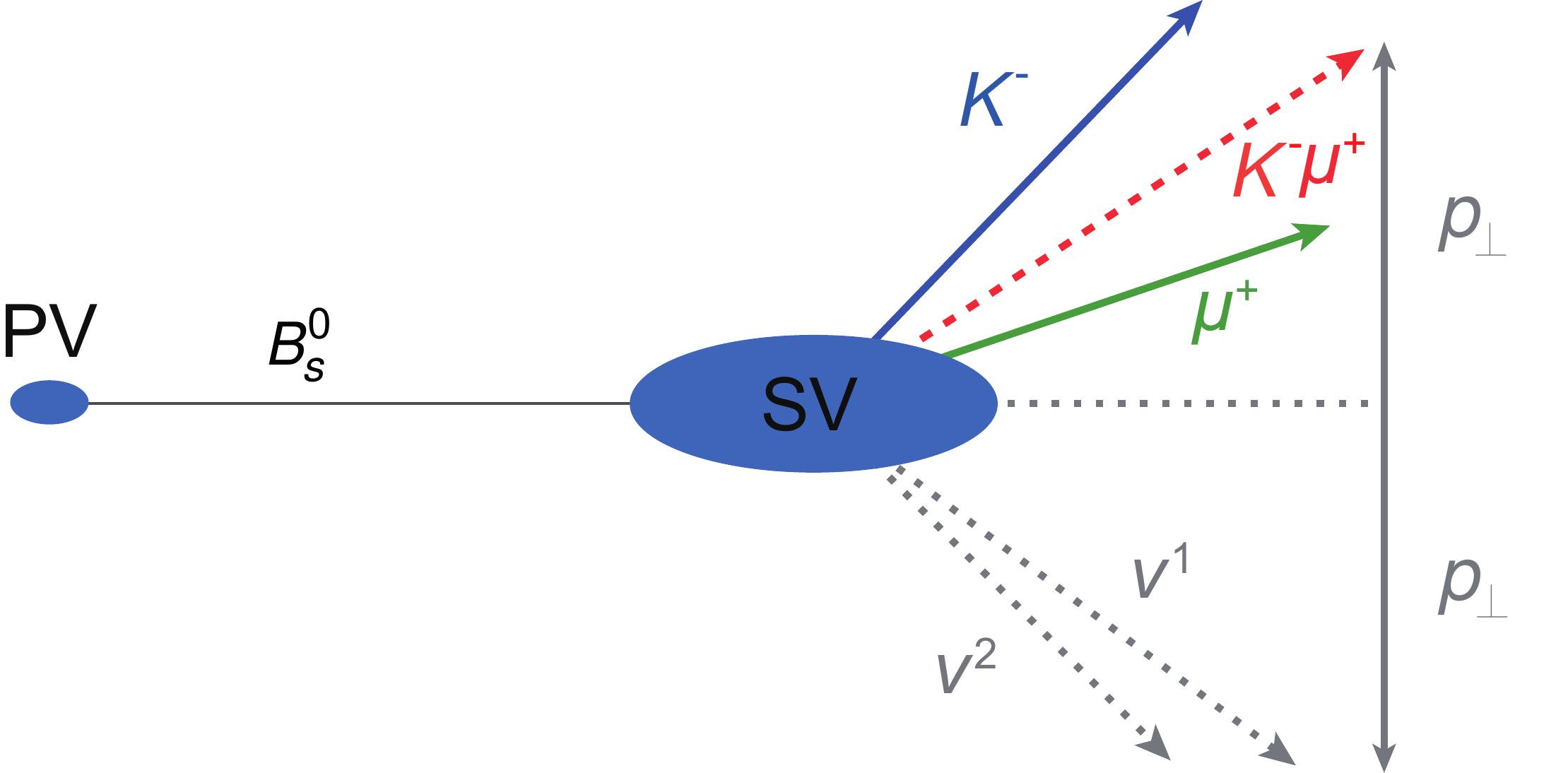}
	\end{minipage}
	\caption{Diagram of conservation of momentum with respect to the  $B_s^0$ flight direction for the decay $B_s^0 \to K\mu\nu$ as an example.}
	\label{topolgy_fig}
\end{figure}

 The decay $B_s^0 \to K\mu\nu$ is used as the example case in this articles and its topology described in Fig.~\ref{topolgy_fig}.

The $B_s^0$ momentum is required to be aligned with the reconstructed flight direction $\vec{F}$~\cite{Stone:2014mza}.
It can be known from the symmetry of the decay that the transverse momentum of the neutrino $p_\perp (\nu)$ must be equal and its sign need to be opposite to the transverse momentum of the visible system $p_\perp$$(K\mu)$ ~\cite{Ciezarek:2016lqu,Aitala:1996vz,LHCb:2020hpv}, that is, shown in the following:
\begin{eqnarray}
	&p_\parallel& = p \cdot \vec{F}, \\
	&p_\perp& = \vert p - p_\parallel \vert = p \times \vec{F}, \\
	&p_\perp (K\mu)& = -p_\perp (\nu).
\end{eqnarray}

\noindent

From the momentum and energy conservation, we then have:
\begin{eqnarray}
	p(B_s^0) &=& p_\parallel(K\mu) + p_\parallel(\nu), \\
	E(B_s^0) &=& E(K\mu) + E(\nu).
\end{eqnarray}
\noindent

Next we use the $B_s^0$ mass constraint to derive $p_\parallel(\nu)$,
\begin{eqnarray*}
	m_{B_s^0}^2 &=& E_{B_s^0}^2 - p_{B_s^0}^2  \\
	&=& E_{K\mu}^2 + 2 \cdot E_{K\mu} \cdot E_\nu + E_\nu^2 -  p_\parallel^2 (K\mu) - p_\parallel^2 (\nu) - 2 \cdot p_\parallel(K\mu) \cdot p_\parallel (\nu) \\
	&=& m_{K\mu}^2 + 2\cdot p_\perp^2(K\mu) + 2 \cdot E_{K\mu} \cdot E_\nu - 2 \cdot p_\parallel(K\mu) \cdot p_\parallel (\nu) .
\end{eqnarray*}
\noindent

Then, we can get a quadratic equation for neutrino momentum in the following form:
\begin{equation}
	\alpha p_\parallel^2 (\nu) + \beta p_\parallel(\nu) + \gamma = 0 ,
	\label{eq:two-fold-solutions}
\end{equation}
where the coefficients are defined as follows:
\begin{eqnarray}
	\alpha &=& 4~[p_\perp^2(K\mu) + m_{K\mu}^2] ,\\
	\beta  &=& 4~p_\parallel(K\mu)~[2~p_\perp^2(K\mu) - m_{B_s^0}^2 + m_{K\mu}^2], \\
	\gamma &=& 4~p_\perp^2(K\mu) [p_\parallel^2 (K\mu) + m_{B_s^0}^2] - [m_{B_s^0}^2 - m_{K\mu}^2]^2.
\end{eqnarray}

Finally, the neutrino momentum parallel to the flight direction can be determined up to a two-fold ambiguity as 
\begin{equation}
	p_\parallel(\nu) = \frac{-\beta \pm \sqrt{\beta^2 - 4\alpha\gamma}}{2 \alpha}.
	\label{eq:last}
\end{equation}
%\textasciitilde

Due to the LHCb detector resolution effects~\cite{Ciezarek:2016lqu}, approximately 20$\sim$40\% of the events selected by the properties of decay chains have an unphysical solution for $p_\parallel(\nu)$, 
that is, the negative values of $\beta^2 - 4\alpha\gamma$.
Such events are discarded in this work.
The $B_s^0$ momentum $p$ and the $q^2$ of signal candidates may now be determined with a two-fold ambiguity.
A choice needs to be made on which of the two solutions of $q^2$ or $p$ will be selected. 
The simplest way is to randomly pick one of the two solutions, but it will lead to a poor resolution of $q^2$ or $p$. 
In order to improve the resolution, a linear regression algorithm is used by using the flight length and the polar angle of the flight vector as the features. 
Based on the above study~\cite{Ciezarek:2016lqu}, in this paper a novel method based on ML has been proposed to further improve the resolution.

\section{Simulation of semileptonic decay production}

The {\tt RapidSim} event generator is used to simulate semileptonic decays in $pp$ collision at $\sqrt{s} = 13$ TeV. About 1 million MC events are generated. 
The paper is using LHCb coordinate system which is defined as $x$ horizontal the beam axis into the LHCb detector, $y$ vertical and $z$ along the beam axis.
Signal heavy-quark hadron events are restricted to be within a pseudorapidity ($\eta$) range $2 < \eta < 5$, which corresponds to the approximate kinematic acceptance of the LHCb detector~\cite{Audurier:2021wqk}.

As the variables used in this study are dependent on the flight direction between the heavy-quark hadron production and its decay vertices, it is necessary for us to model the resolution in associated features, that is, we need to apply a proper smearing at first in order to simulate expected experimental resolution.
The $x$ and $y$ coordinates of the heavy-quark hadron decay vertices are smeared by a Gaussian distribution with a sigma value of $\rm \pm 20~\mu m$. 
A much larger resolution of $\rm \pm 200~\mu m$ is applied in the $z$ direction~\cite{Ciezarek:2016lqu}.
To reflect the known performance from the LHCb VELO detector~\cite{Geertsema:2020pmc,Ciezarek:2016lqu}, the resolutions of production vertices for $x$, $y$ and $z$ ordinates are assumed at $\rm \pm 13~\mu m$, $\rm \pm 13~\mu m$ and $\rm \pm 70~\mu m$, respectively.   
In all presented studies, the smeared flight length needs to be larger than 3~mm. 
These assumptions approximately meet the effect of online and offline selection from heavy-quark hadron decays in LHCb~\cite{Aaij:2018jht,Ciezarek:2016lqu}. 

\section{features and regressors}
\label{sec:inputVars}
The regression analysis is a set of statistical methods used for estimating the targeted value based on the relationships between regressor and features~\cite{Pedregosa:2011ork}.
Therefore it is important to select well suited regressors and efficient features for different user-case scenarios. 

In Ref.~\cite{Ciezarek:2016lqu}, the momentum of the $b$-hadron as the mother particle is inferred based on
a linear regression algorithm using two flight variables, $1/\sin(\theta_{\rm flight})$ and $\left| \vec{\it F} \right|$, 
where %$\beta_{0,1,2}$ are parameters to be determined by the regression analysis,
$\left| \vec{F} \right|$ represents the flight distance of mother particle and
$\theta_{\rm flight}$ is the polar angle of the flight vector.
In our case, five sets of features have been chosen, as summarised in Table~\ref{tab:inputVars}.
All features are selected based on %Equation~\ref{eq:p_flight} and 
Section~\ref{sec:theorecticalDerivation} and those used in Ref.~\cite{Ciezarek:2016lqu}, 
where $F_x$, $F_y$, $F_z$ are the components of $\vec{F}$.
Three different regressors
are studied in this paper, labeled as “\mbox{Regressor A-C}"~\cite{Garcia:2021znz,Lebese:2021foi,Yuksel:2021nae}, shown in Table~\ref{tab:inputVars}.
These regressors are selected from a full range of regression models included in the {\tt scikit-learn} toolkit. 

\begin{table}[htp]
	\centering
	\caption{Sets of Features and Regressors used in this study.}
	\vspace{0.2cm}
	\label{tab:inputVars}
	\begin{tabular}{ccc} \hline\hline
		Description & Features & Regressor \\ \hline
		Label A & $|\vec{F}|$ and $\rm 1/sin(\theta_{\rm flight})$ & - \\
		Label B & $F_x$, $F_y$, $F_z$ & - \\
		Label C & $F_x$, $F_y$, $F_z$ and $\rm 1/sin(\theta_{\rm flight})$ & - \\
		Label D & Label A + $p_\parallel(K\mu)$ and $p_\perp^2(K\mu)$ & - \\
		Label E & Label C + $p_\parallel(K\mu)$ and $p_\perp^2(K\mu)$ & -\\
		Regressor A & - & Linear Regressor \\
		Regressor B & - & GradientBoosting Regressor \\%GB Regressor \\
		Regressor C & - & MLP Regressor \\ \hline
	\end{tabular}
\end{table}

\begin{figure*}[htbp]
	\centering
	\includegraphics[width=.8\textwidth,clip,trim=0 0 0 0]{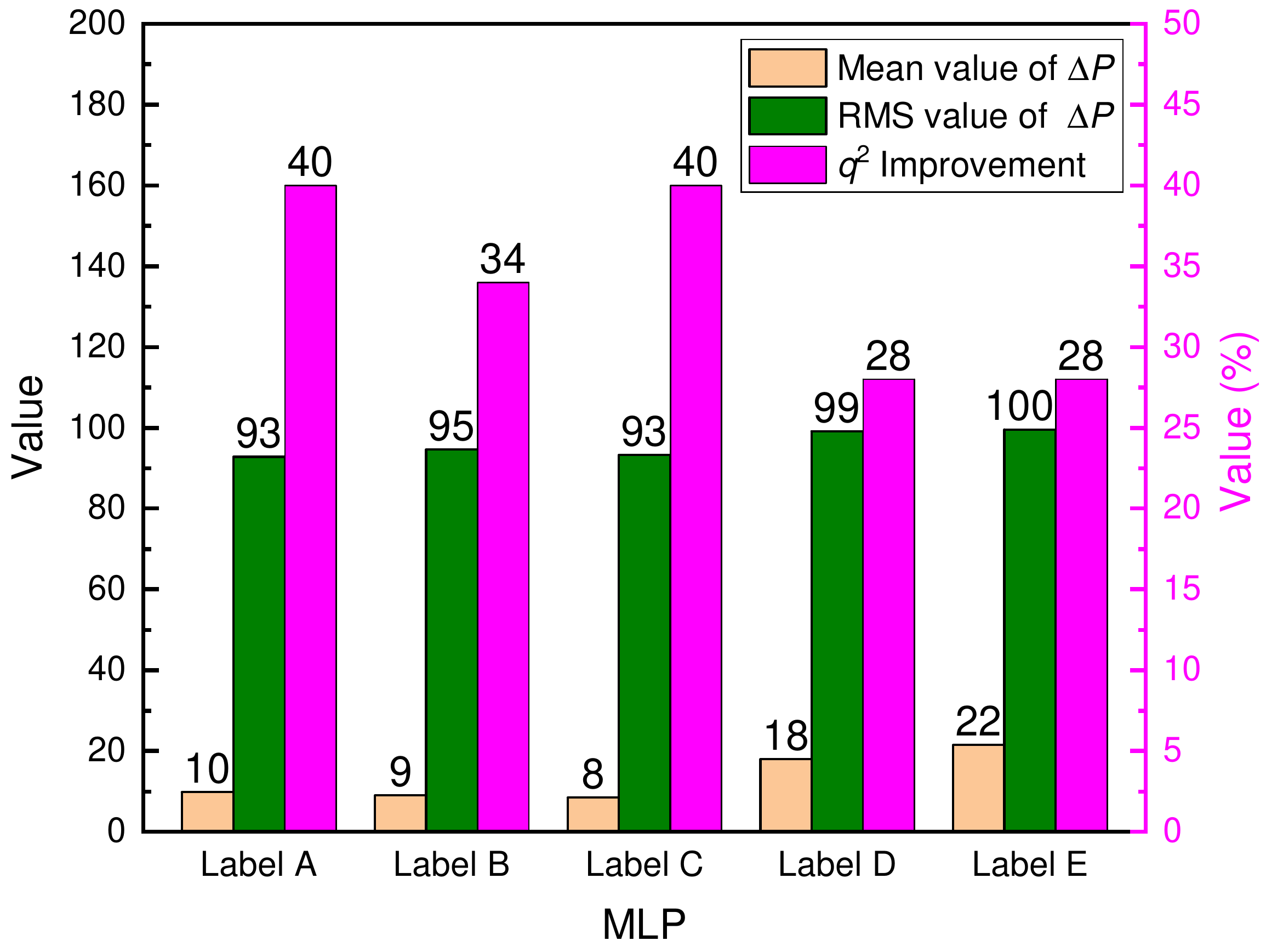}
	\caption{The performance of different sets of features with MLP regressor.
	 %$\Delta P = P_{best} - P_{true}$;
	“Label A": $|\vec{F}|$ and $\rm 1/sin(\theta_{\rm flight})$;
	“Label B": $F_x$, $F_y$, $F_z$;
	“Label C": $F_x$, $F_y$, $F_z$ and $\rm 1/sin(\theta_{\rm flight})$;
	“Label D": “Label A" + $p_\parallel(K\mu)$ and $p_\perp^2(K\mu)$;
	“Label E": “Label C" + $p_\parallel(K\mu)$ and $p_\perp^2(K\mu)$.
	}
	\label{fig:4_vars}
\end{figure*}

\begin{figure}[htbp]
          \centering
          \includegraphics[width=8cm]{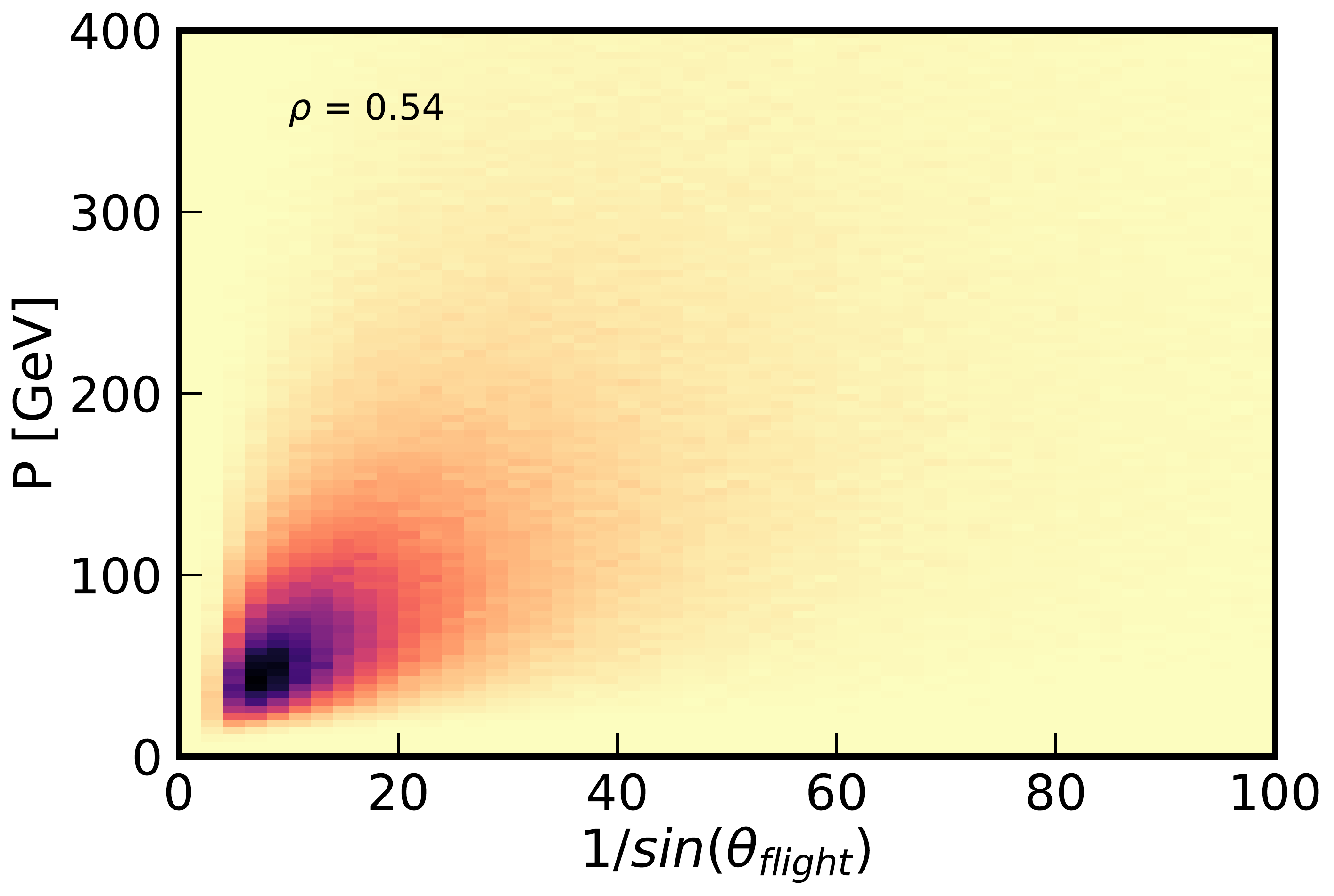}
          \includegraphics[width=8cm]{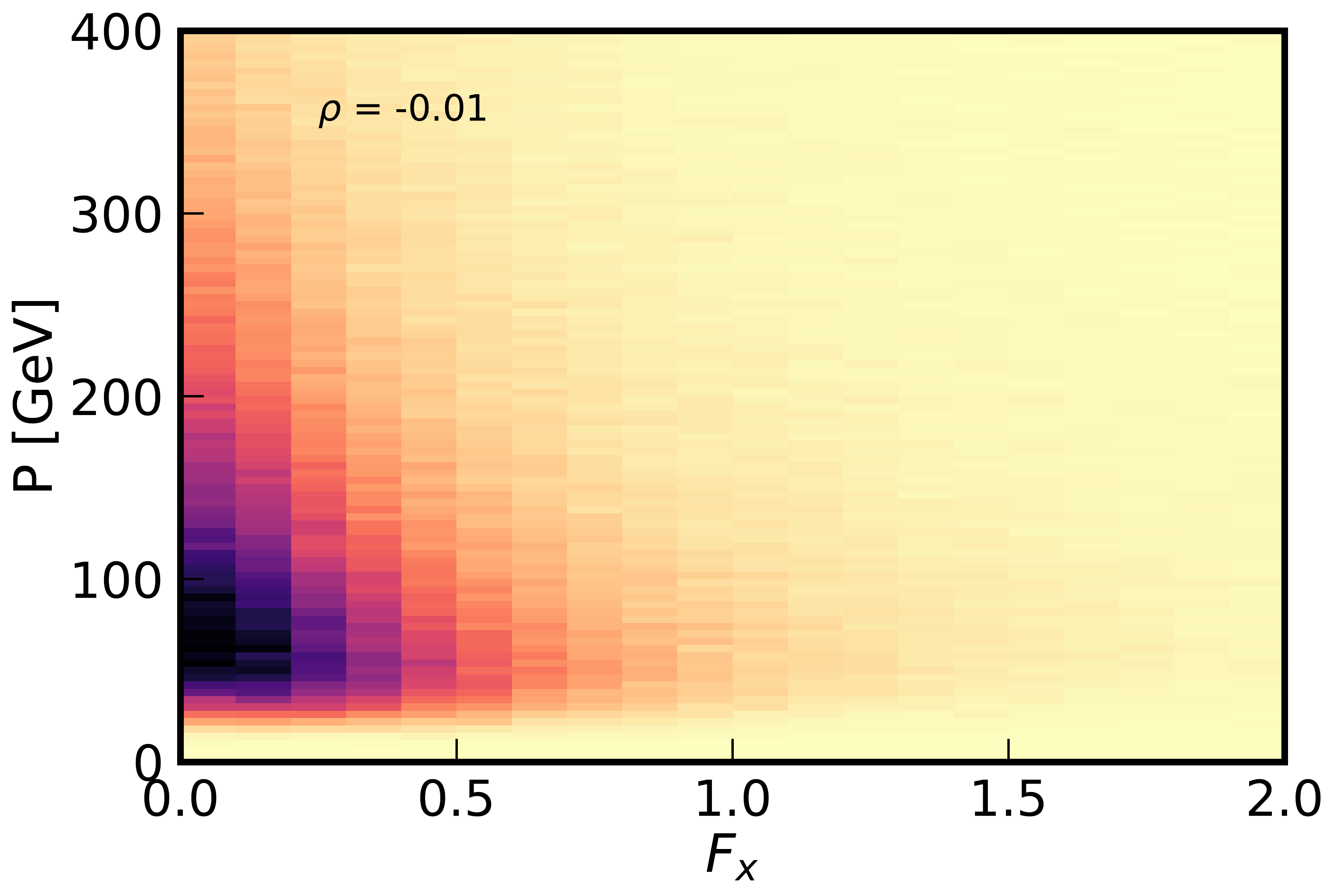}
          \includegraphics[width=8cm]{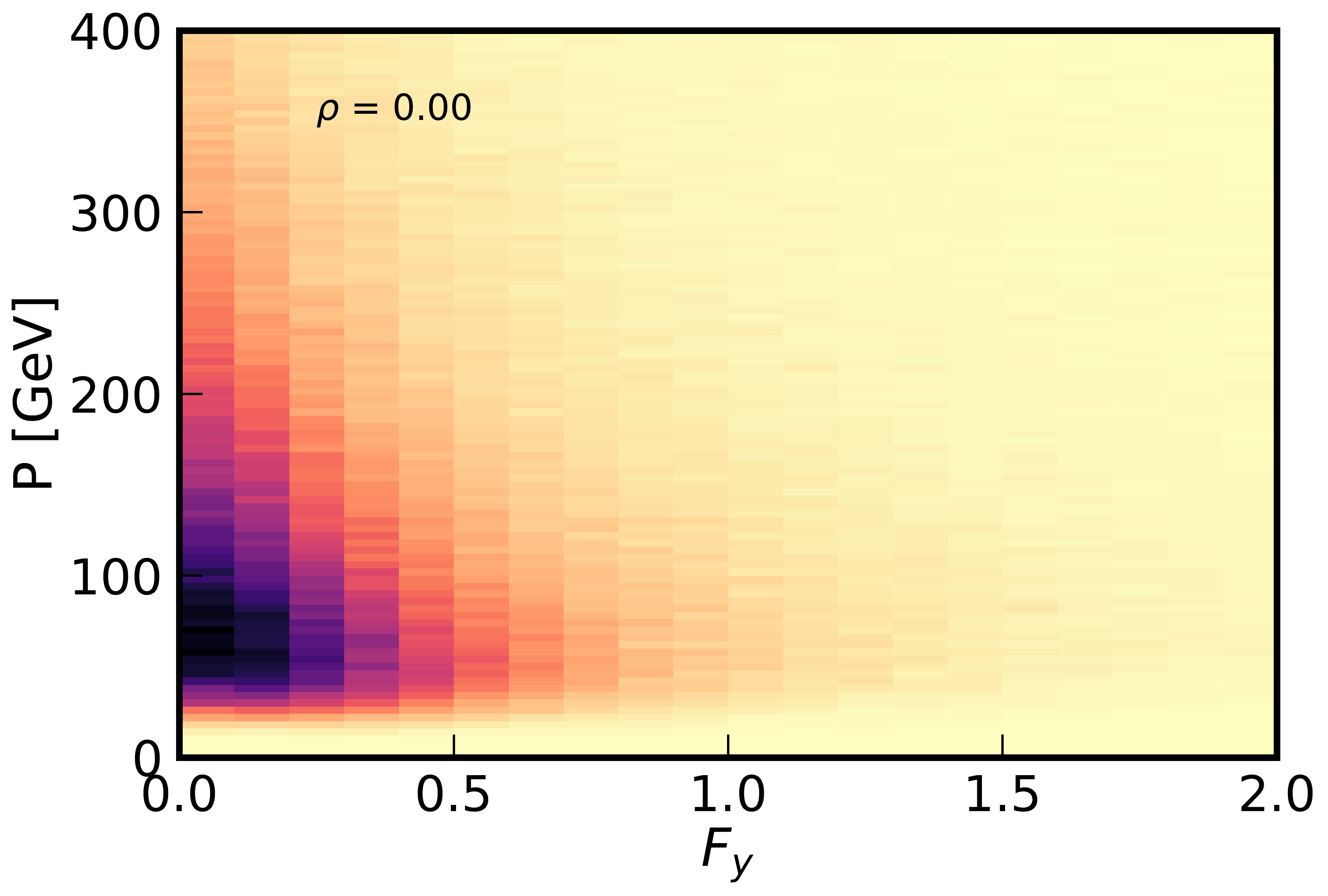} 
          \includegraphics[width=8cm]{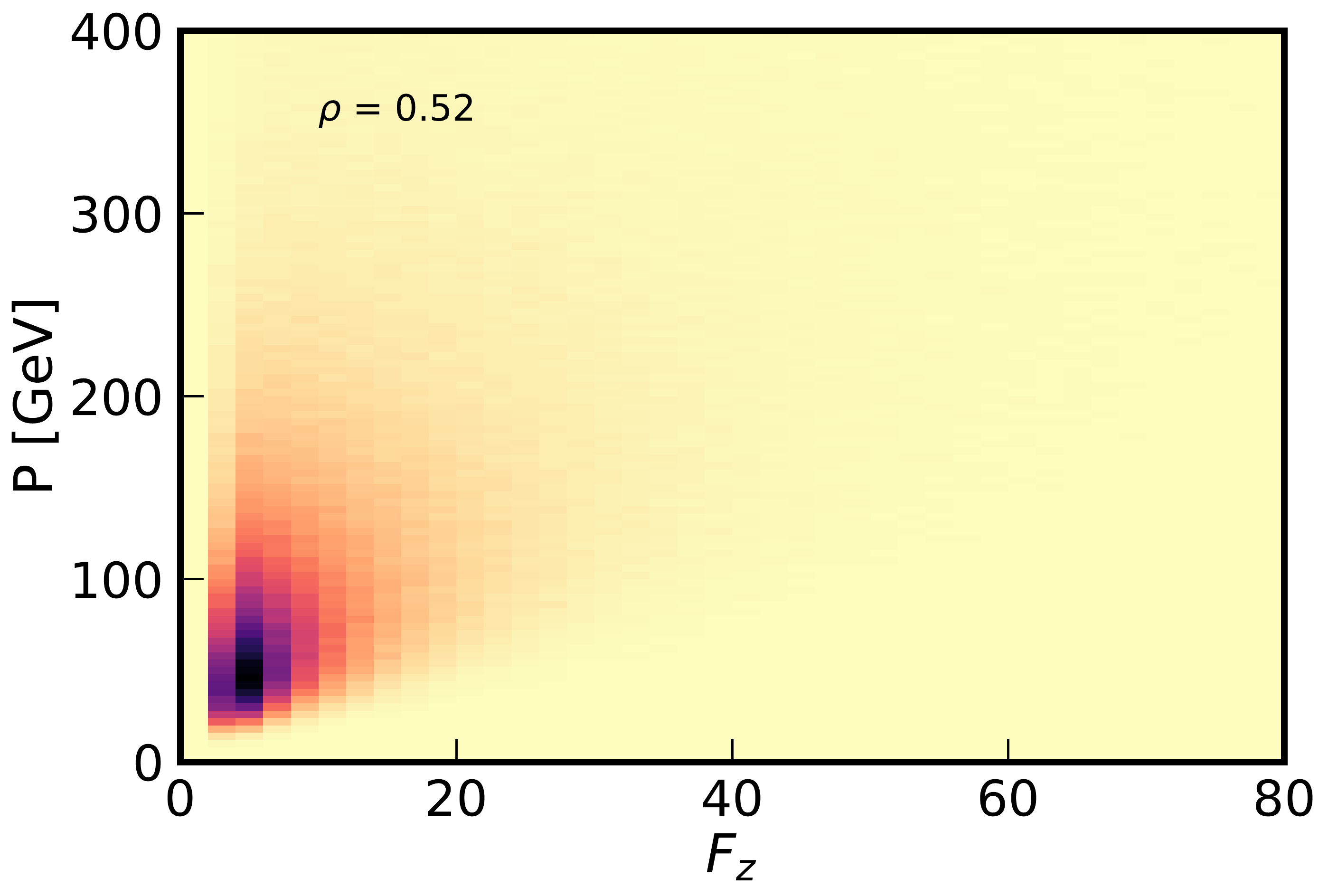}
    \caption{The distribution of $\rm 1/sin(\theta_{\rm flight})$, $F_x$, $F_y$, and $F_z$ versus the $b$-hadron momentum.}
          \label{sinthetap}
\end{figure}

To test the performance of different sets of features and select the best one, we make conditional experiments.
Figure~\ref{fig:4_vars} shows the performance on $q^2$ improvement 
and the Root Mean Square (RMS) value of reconstructed $b$-hadron momentum resolution ($\Delta P \equiv P_{best} - P_{true}$) with different sets of input variables 
from the MLP regressor. 
It indicates “Label A" and “Label C" have the same performance on $q^2$ improvement, which increased by 40\%, while other sets are less than 35\%. 
The mean and RMS values of $\Delta P$ in 
“Label A", “Label B" and “Label C" are (10, 93) MeV/$c$, (9, 95) MeV/$c$, and (8, 93) MeV/$c$, respectively.
Based on the obtained results, we select “Label C" as the main method for this study, that is, $F_x$, $F_y$, $F_z$ and $\rm 1/sin(\theta_{\rm flight})$. 
Figure~\ref{sinthetap} shows the distributions of $\rm 1/sin(\theta_{\rm flight})$, $F_x$, $F_y$, and $F_z$ versus the $b$-hadron momentum with the correlation coefficients of 0.54, -0.01, -0.00, and 0.52.

\begin{figure*}[htbp]
	\centering
	\includegraphics[width=.8\textwidth,clip,trim=0 0 0 0]{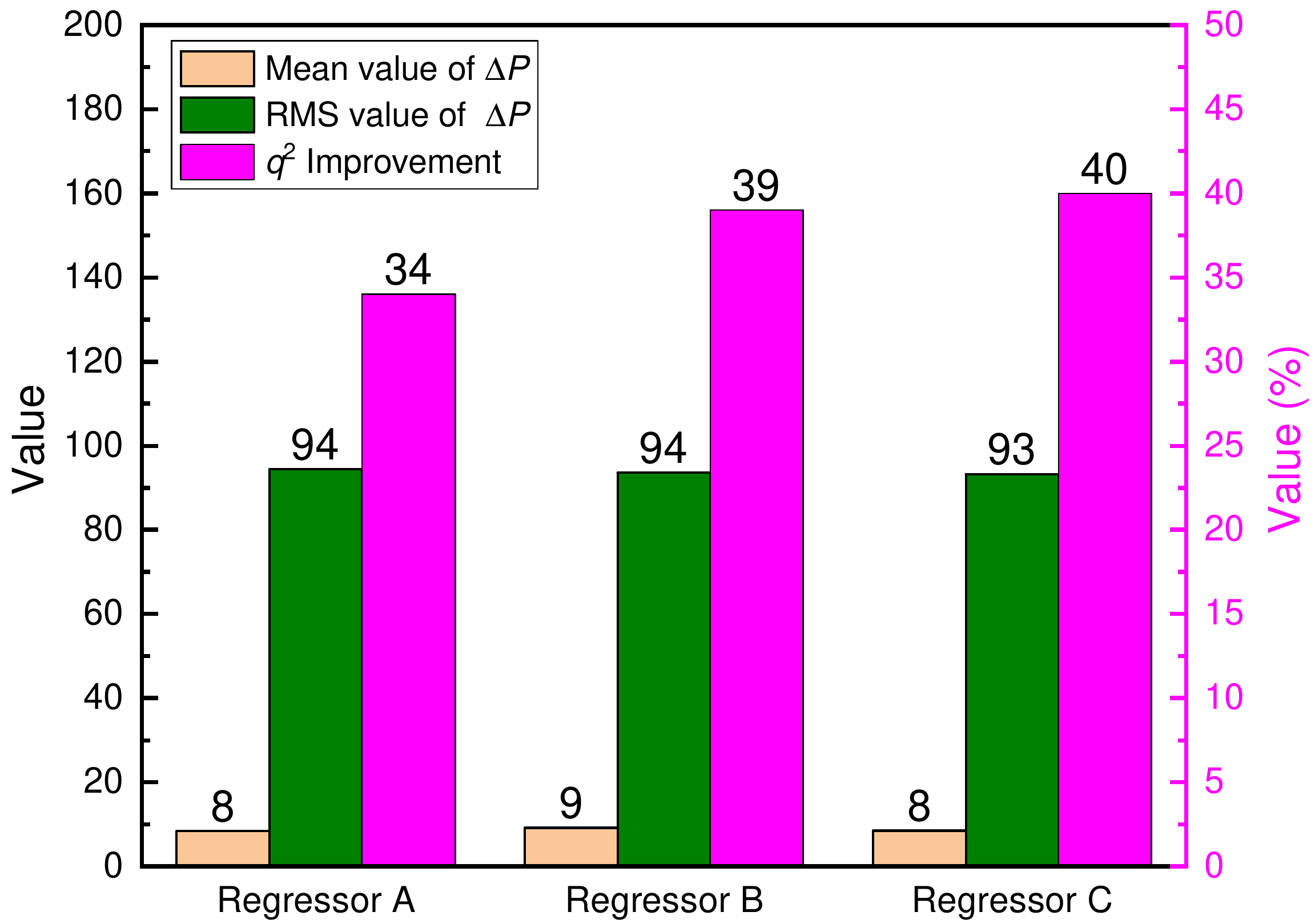}
	\caption{The performance of different regressors with “Label C" variables.
	 $\Delta P = P_{best} - P_{true}$;
	“Regressor A": Linear Regressor;
	“Regressor B": GB Regressor;
	“Regressor C": MLP Regressor.
	}
	\label{fig:mlp_performance}
\end{figure*}

Once the input features are determined, the best regressor is selected by a similar method.
Figure~\ref{fig:mlp_performance} shows the performance on $q^2$ improvement and the RMS value of 
 momentum using the different regressors based on the “Label C" input features.
The $q^2$ resolution increase has been observed for Regressor A, B and C as 34\%, 39\% and 40\%, 
while the (mean and RMS) values of $\Delta P$ for that are (8, 94) MeV/$c$, (9, 94) MeV/$c$, and (8, 93) MeV/$c$, respectively.
The best of features is “Label C" which consists of $F_x$, $F_y$, $F_z$ and $\rm 1/sin(\theta_{\rm flight})$, while the best regressor is the MLP regressor.

\section{Performance of MLP regressor}
This section describes the applications of the best regressor, MLP regressor, for different semileptonic decays, such as $B_s^0 \to K\mu\nu$, $B_s^0 \to D_s\mu\nu$, $\Lambda_b^0 \to p\mu\nu$ and $\Lambda_b^0 \to \Lambda_c\mu\nu$. 

\subsection{Tests on $B_s^0 \to K\mu\nu$ channel}
\label{subsec:TestOnBs2kmunu}
\begin{figure*}[htbp]
	\centering
	\includegraphics[width=.8\textwidth,clip,trim=0 0 0 0]{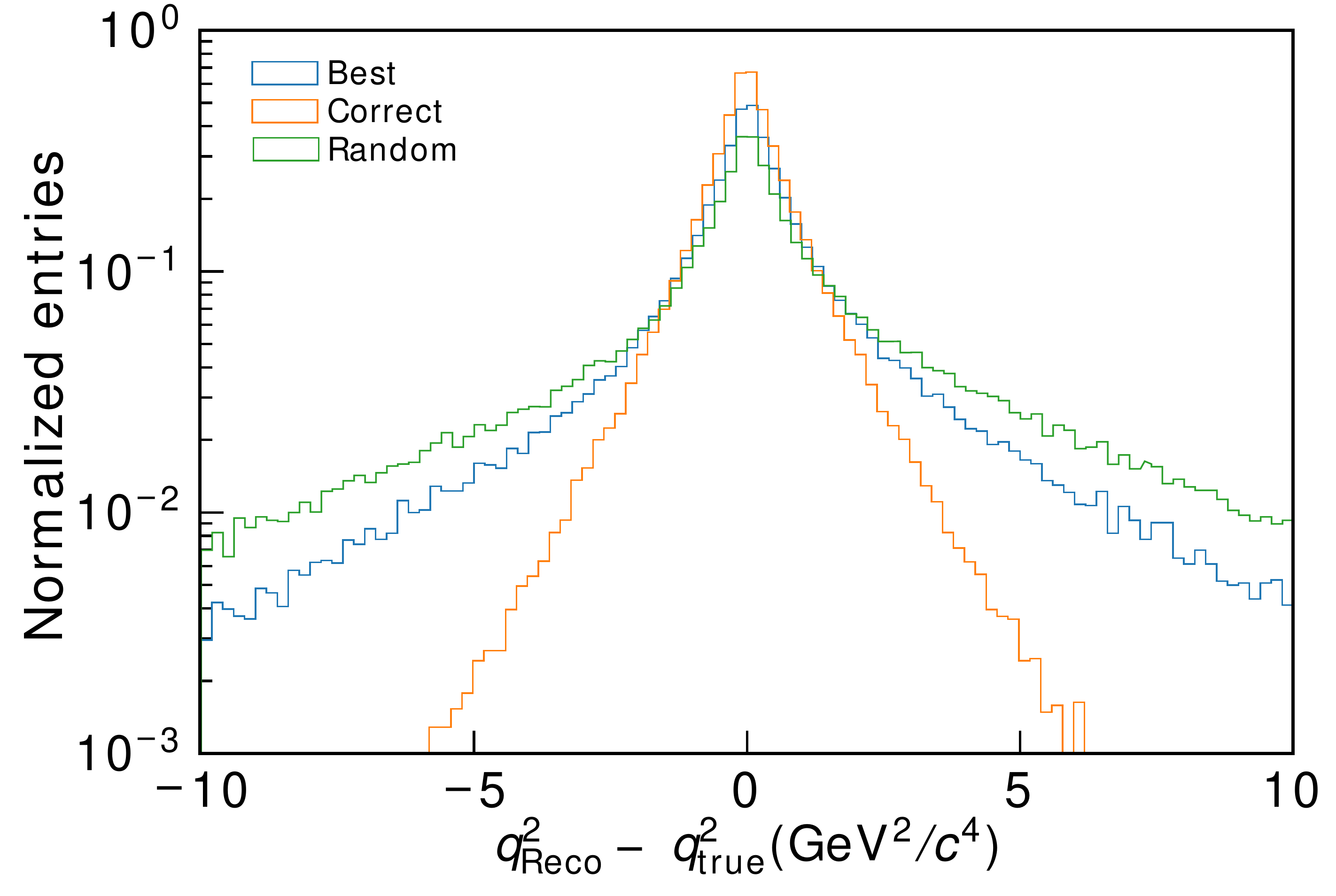}
	\caption{Comparison of $q^2$ resolution in different conditions with MLP regressor and “Label C" feature.}
	\label{Bs2kmunu_figs}
\end{figure*}

\begin{figure}[ht!]
          \centering
          \includegraphics[width=.8\textwidth,clip,trim=0 0 0 0]{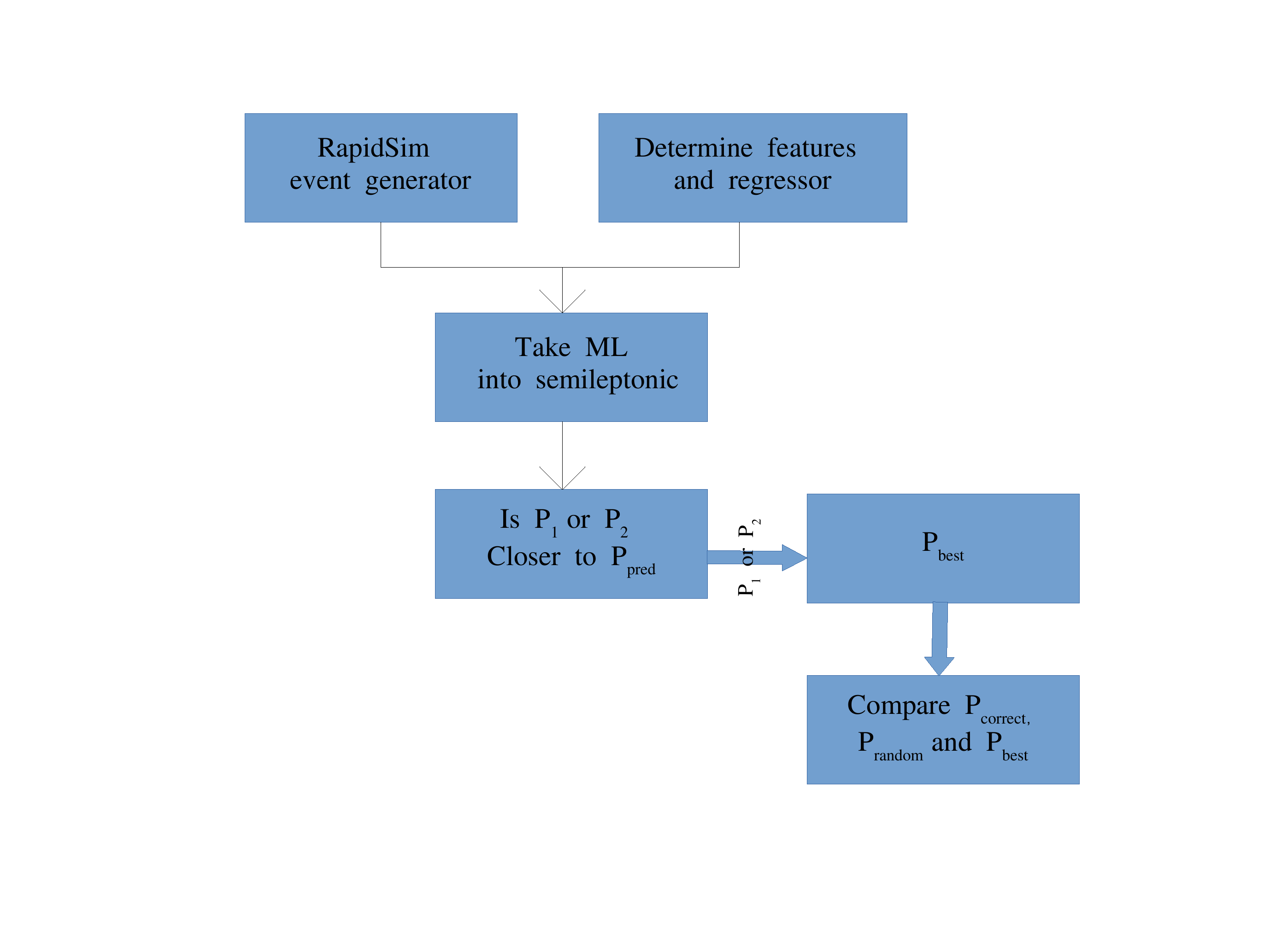}
    \caption{The flowchart of the methodology.}
          \label{flowchart}
\end{figure}

$B_s^0 \to K\mu\nu$ decay channel has been used to study the improvement of $q^2$ resolution with MLP regressor and “Label C" feature.
Figure~\ref{Bs2kmunu_figs} shows the distributions of $q^2$ resolution ($\Delta q^2\equiv q^2_{Reco} - q^2_{true}$, where $q^2_{Reco}$ and $q^2_{true}$ are the reconstructed and input $q^2$ value, respectively) in different conditions, labeled as “Best", “Correct" and “Random".
“Best" represents the result which corresponds to the regression value.
“Correct" is defined as the solution being the one closest to the true $q^2$ from the input MC. The value is set up here for comparison. 
“Random" is the solution based on selecting a random result of Eq.~\ref{eq:last}. 
The result indicates an obvious improvement from “Best" compared with that from ``Random''.
The flowchart of the methodology is shown in Fig.~\ref{flowchart}.

Table~\ref{tab:rms_and_improvements} shows the resolution on the reconstructed $q^2$ for different ways of selecting a solution of the two-fold ambiguity, and shows the improvement on the resolution of reconstructed $q^2$ compared to a random selection.
Using the output of the MLP regression model with “Label C" feature to select a solution improves the resolution on the reconstructed $q^2$ by $\sim$40\% when compared with a random selection. 

\begin{table}[htp]
	\centering
	\caption{Resolution on reconstructed $q^2$ after selecting one of two solutions, and improvements on the resolution of reconstructed $q^2$ compared to a random selection.}
	\vspace{0.2cm}
	\label{tab:rms_and_improvements}
	\begin{tabular}{ccc} \hline\hline
		Solution & RMS ($\rm GeV^2/c^4$) & Improvement (\%) \\ \hline
		Correct  & 1.2 & - \\
		Best & 3.02 & 40\% \\
		Random & 4.23 & - \\ \hline
	\end{tabular}
\end{table}

\begin{figure} % figuur 1
	\begin{minipage}{\columnwidth}
		\centering
		\includegraphics[width=.8\textwidth,clip,trim=0 0 0 0]{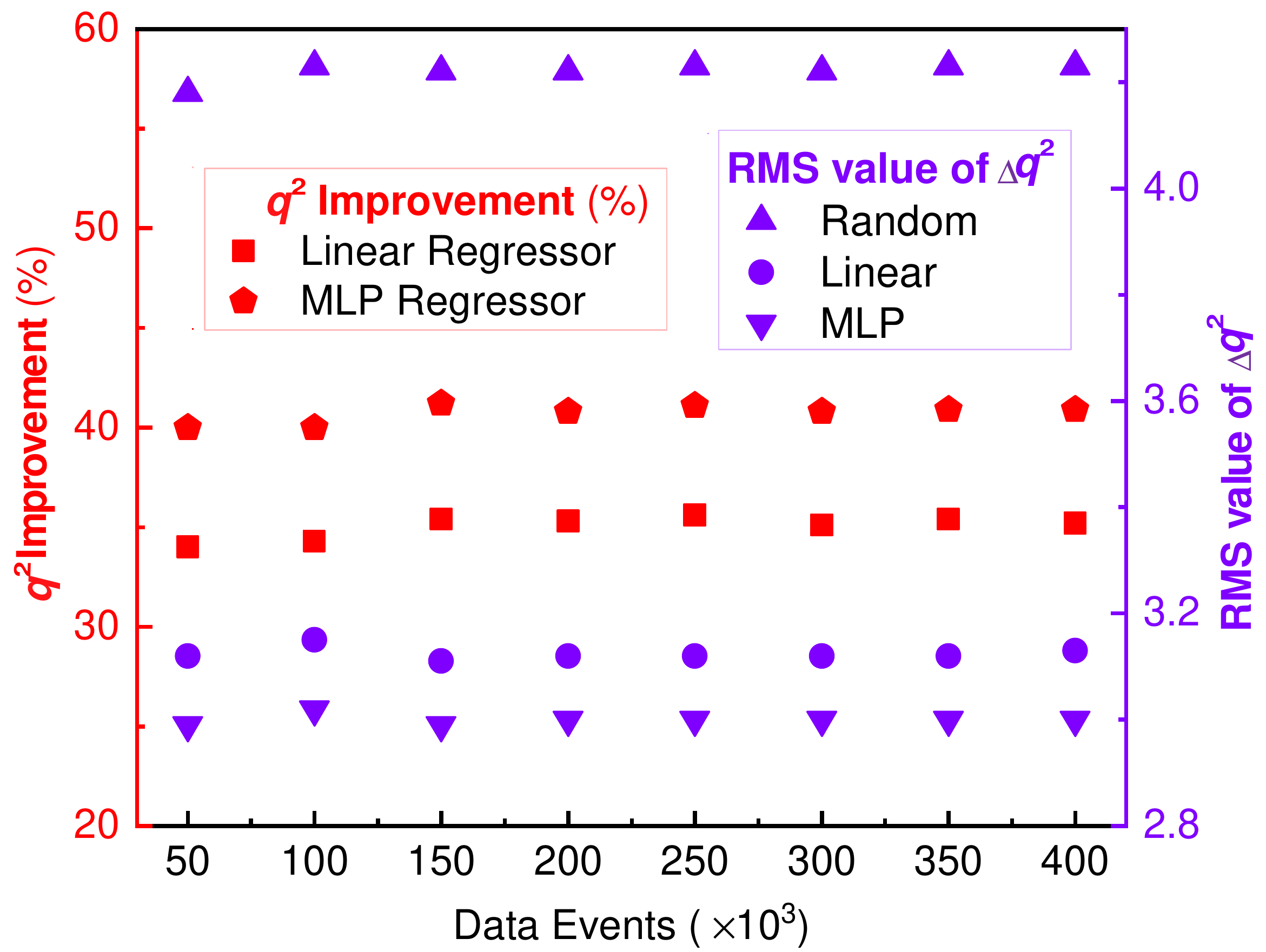}
	\end{minipage}
	\caption{The improvement of reconstructed $q^2$ resolution and the RMS value of $\Delta q^2$ with different numbers of data events.}
	\label{fig:Bs2kmunu_test_size}
\end{figure}

To illustrate the robustness of the model, data samples with different sizes are tested. Figure~\ref{fig:Bs2kmunu_test_size} shows the improvement of reconstructed $q^2$ resolution and the RMS value of $\Delta q^2$ based on various input statistics, with the linear regressor included for comparison. 
The improvement of $q^2$ resolution in case of the MLP regressor is on average higher by $5\%$ with respect to values obtained using the linear regressor.
The RMS values of $\Delta q^2$ from the MLP regressor are clearer smaller than those from the Random Choice in all tested data samples, meanwhile on average around $40\%$ of improvements for reconstructed $q^2$ resolution can be achieved by MLP regressor. 

To summarize this part, the MLP regression method can significantly improve the $q^2$ resolution up to $40\%$ when compared to the random choice or up to $5\%$ when compared with the linear regressor, 
so that a more precise measurement on the ratio $\rm \frac{\vert V_{ub} \vert}{\vert V_{cb} \vert}$  based on our method is expected in the $B_s^0 \to K\mu\nu$ channel. 

\subsection{Tests on other channels} 
In order to scrutinize obtained results, selected method is applied to other semileptonic decays and revaluated.
Three channels, namely, $B_s^0 \to D_s\mu\nu$, $\Lambda_b^0 \to p\mu\nu$ and $\Lambda_b^0 \to \Lambda_c\mu\nu$ have been chosen to check the performance.
The performance tests on other channels confirm that using the output of MLP regression, improved $q^2$ resolution can be obtained in all tested channels.  
More specially, the resolution improvement on the reconstructed $q^2$ with respect to a random selection is, on average, 40\% by using the MLP regressor in the $B_s^0 \to D_s\mu\nu$ decay mode. % from 30\% using linear regression.
For the channels of $\Lambda_b^0 \to p\mu\nu$ and $\Lambda_b^0 \to \Lambda_c\mu\nu$, the resolution is improved by 37\% and 20\% on average, respectively. % from 27\% and 15\%.
The MLP regressor, when compared with the linear regressor, can on average result in $\sim 5\%$ improvement on the obtained $q^2$ resolution for all studied decay channels. 

\section{Conclusions}
A novel method to improve the $q^2$ resolution in semileptonic decays using a ML approach is studied in this paper. 
The information of flight vector ($\rm F_x$, $\rm F_y$, $\rm F_z$ and  $\rm 1/sin(\theta_{\rm flight})$), labeled as “Label C", shows the highest discrimination power, while the MLP regressor is the best regressor. 
We found:
\begin{itemize}
   \item Using the MLP regression model with “Label C" feature improves the resolution on the reconstructed $q^2$ by an average of $\sim40\%$ when compared to the random choice or up to $5\%$ when compared with the linear regressor method introduced in Ref.~\cite{Ciezarek:2016lqu}, when the decay $B_s^0 \to K\mu\nu$ is used as a test channel.
   \item The method also have similar performance on improving the reconstructed $q^2$ resolution in a wide range of semileptonic decays, namely $B_s^0 \to D_s\mu\nu$, $\Lambda_b^0 \to p\mu\nu$ and $\Lambda_b^0 \to \Lambda_c\mu\nu$.
    \item What's more, the proposed method method can potentially improve measurements of differential decay rates of semileptonic heavy flavour hadrons decays in hadron collider experiments such as LHCb.
    \item The studies presented here use the example of the LHCb experiment, but the ideas should be available to any other hadron collider experiment in the current and future.
\end{itemize}
However, the room for improvement using sole software means is rather limited due to the experimental resolution of the vertex positioning that we have assumed ($\pm 200 \mu$m in the $z$ direction, $\pm 20 \mu$m in $x$ or $y$ direction) based on the LHCb experiment.

\section*{Acknowledgement}

This work was supported by grants from Natural Science Foundation of China (no.11735010, U1932108, U2032102, 12061131006). 
The authors would like to thank Murphy Zheng (Murphy-Zheng Creative Studio) for polishing Figure~\ref{topolgy_fig} and Zhihao Xu (University of Chinese Academy of Sciences) for useful discussion. M.S. acknowledges support from the European Union's Horizon 2020 research and innovation programme under grant agreement No. 714536: PRECISION.

	
\end{document}